\documentclass[iop]{emulateapj}
\usepackage{aas_macros}

\usepackage{color}
\usepackage[dvipsnames,svgnames]{xcolor}

\usepackage{natbib}

\usepackage{amsmath}
\usepackage{xspace}

\mathchardef\mhyphen="2D

\newlength{\dhatheight}

\newcommand{\code}[1]{\texttt{#1}\xspace}

\providecommand\physrep{\ref@jnl{Phys.~Rep.}}%
\providecommand\apjs{\ref@jnl{ApJS}}%
\providecommand{\jcap}{\ref@jnl{JCAP}}%

\newcommand{\lsim}{\lower0.6ex\vbox{\hbox{$ \buildrel{\textstyle <}\over{\sim}\ $}}}

\shorttitle{Sixth star cluster in the Fornax dwarf spheroidal galaxy}
\shortauthors{Wang et~al.}
\begin{document}

\title{Rediscovery of the sixth star cluster in the Fornax dwarf spheroidal galaxy}

\author{
M.~Y.~Wang\altaffilmark{1},
S.~Koposov\altaffilmark{1,2},
A.~Drlica-Wagner\altaffilmark{3,4},
A.~Pieres\altaffilmark{5,6},
T.~S.~Li\altaffilmark{3,4},
T.~de Boer\altaffilmark{2},
K.~Bechtol\altaffilmark{7,8},
V.~Belokurov\altaffilmark{2},
A. B.~Pace\altaffilmark{9},
T.~M.~C.~Abbott\altaffilmark{10},
J.~Annis\altaffilmark{3},
E.~Bertin\altaffilmark{11,12},
D.~Brooks\altaffilmark{13},
E.~Buckley-Geer\altaffilmark{3},
D.~L.~Burke\altaffilmark{14,15},
A.~Carnero~Rosell\altaffilmark{16,6},
M.~Carrasco~Kind\altaffilmark{17,18},
J.~Carretero\altaffilmark{19},
L.~N.~da Costa\altaffilmark{6,20},
J.~De~Vicente\altaffilmark{16},
S.~Desai\altaffilmark{21},
H.~T.~Diehl\altaffilmark{3},
P.~Doel\altaffilmark{13},
J.~Estrada\altaffilmark{3},
B.~Flaugher\altaffilmark{3},
P.~Fosalba\altaffilmark{22,23},
J.~Frieman\altaffilmark{3,4},
J.~Garc\'ia-Bellido\altaffilmark{24},
D.~W.~Gerdes\altaffilmark{25,26},
D.~Gruen\altaffilmark{27,14,15},
R.~A.~Gruendl\altaffilmark{17,18},
J.~Gschwend\altaffilmark{6,20},
G.~Gutierrez\altaffilmark{3},
D.~L.~Hollowood\altaffilmark{28},
K.~Honscheid\altaffilmark{29,30},
B.~Hoyle\altaffilmark{31,32},
D.~J.~James\altaffilmark{33},
S.~Kent\altaffilmark{3,4},
K.~Kuehn\altaffilmark{34},
N.~Kuropatkin\altaffilmark{3},
M.~A.~G.~Maia\altaffilmark{6,20},
J.~L.~Marshall\altaffilmark{9},
F.~Menanteau\altaffilmark{17,18},
R.~Miquel\altaffilmark{35,19},
A.~A.~Plazas\altaffilmark{36},
E.~Sanchez\altaffilmark{16},
B.~Santiago\altaffilmark{37,6},
V.~Scarpine\altaffilmark{3},
R.~Schindler\altaffilmark{15},
M.~Schubnell\altaffilmark{26},
S.~Serrano\altaffilmark{22,23},
I.~Sevilla-Noarbe\altaffilmark{16},
M.~Smith\altaffilmark{38},
R.~C.~Smith\altaffilmark{10},
F.~Sobreira\altaffilmark{39,6},
E.~Suchyta\altaffilmark{40},
M.~E.~C.~Swanson\altaffilmark{18},
G.~Tarle\altaffilmark{26},
D.~Thomas\altaffilmark{41},
D.~L.~Tucker\altaffilmark{3},
and A.~R.~Walker\altaffilmark{10}
\\ \vspace{0.2cm} (DES Collaboration) \\
}

\affil{$^{1}$ Department of Physics, Carnegie Mellon University, Pittsburgh, Pennsylvania 15312, USA}
\affil{$^{2}$ Institute of Astronomy, University of Cambridge, Madingley Road, Cambridge CB3 0HA, UK}
\affil{$^{3}$ Fermi National Accelerator Laboratory, P. O. Box 500, Batavia, IL 60510, USA}
\affil{$^{4}$ Kavli Institute for Cosmological Physics, University of Chicago, Chicago, IL 60637, USA}
\affil{$^{5}$ Observat\'orio Nacional, Rua General Jos\'e Cristino, 77, Rio de Janeiro, RJ, 20921-400, Brazil}
\affil{$^{6}$ Laborat\'orio Interinstitucional de e-Astronomia - LIneA, Rua Gal. Jos\'e Cristino 77, Rio de Janeiro, RJ - 20921-400, Brazil}
\affil{$^{7}$ LSST, 933 North Cherry Avenue, Tucson, AZ 85721, USA}
\affil{$^{8}$ Physics Department, 2320 Chamberlin Hall, University of Wisconsin-Madison, 1150 University Avenue Madison, WI  53706-1390}
\affil{$^{9}$ George P. and Cynthia Woods Mitchell Institute for Fundamental Physics and Astronomy, and Department of Physics and Astronomy, Texas A\&M University, College Station, TX 77843,  USA}
\affil{$^{10}$ Cerro Tololo Inter-American Observatory, National Optical Astronomy Observatory, Casilla 603, La Serena, Chile}
\affil{$^{11}$ CNRS, UMR 7095, Institut d'Astrophysique de Paris, F-75014, Paris, France}
\affil{$^{12}$ Sorbonne Universit\'es, UPMC Univ Paris 06, UMR 7095, Institut d'Astrophysique de Paris, F-75014, Paris, France}
\affil{$^{13}$ Department of Physics \& Astronomy, University College London, Gower Street, London, WC1E 6BT, UK}
\affil{$^{14}$ Kavli Institute for Particle Astrophysics \& Cosmology, P. O. Box 2450, Stanford University, Stanford, CA 94305, USA}
\affil{$^{15}$ SLAC National Accelerator Laboratory, Menlo Park, CA 94025, USA}
\affil{$^{16}$ Centro de Investigaciones Energ\'eticas, Medioambientales y Tecnol\'ogicas (CIEMAT), Madrid, Spain}
\affil{$^{17}$ Department of Astronomy, University of Illinois at Urbana-Champaign, 1002 W. Green Street, Urbana, IL 61801, USA}
\affil{$^{18}$ National Center for Supercomputing Applications, 1205 West Clark St., Urbana, IL 61801, USA}
\affil{$^{19}$ Institut de F\'{\i}sica d'Altes Energies (IFAE), The Barcelona Institute of Science and Technology, Campus UAB, 08193 Bellaterra (Barcelona) Spain}
\affil{$^{20}$ Observat\'orio Nacional, Rua Gal. Jos\'e Cristino 77, Rio de Janeiro, RJ - 20921-400, Brazil}
\affil{$^{21}$ Department of Physics, IIT Hyderabad, Kandi, Telangana 502285, India}
\affil{$^{22}$ Institut d'Estudis Espacials de Catalunya (IEEC), 08034 Barcelona, Spain}
\affil{$^{23}$ Institute of Space Sciences (ICE, CSIC),  Campus UAB, Carrer de Can Magrans, s/n,  08193 Barcelona, Spain}
\affil{$^{24}$ Instituto de Fisica Teorica UAM/CSIC, Universidad Autonoma de Madrid, 28049 Madrid, Spain}
\affil{$^{25}$ Department of Astronomy, University of Michigan, Ann Arbor, MI 48109, USA}
\affil{$^{26}$ Department of Physics, University of Michigan, Ann Arbor, MI 48109, USA}
\affil{$^{27}$ Department of Physics, Stanford University, 382 Via Pueblo Mall, Stanford, CA 94305, USA}
\affil{$^{28}$ Santa Cruz Institute for Particle Physics, Santa Cruz, CA 95064, USA}
\affil{$^{29}$ Center for Cosmology and Astro-Particle Physics, The Ohio State University, Columbus, OH 43210, USA}
\affil{$^{30}$ Department of Physics, The Ohio State University, Columbus, OH 43210, USA}
\affil{$^{31}$ Max Planck Institute for Extraterrestrial Physics, Giessenbachstrasse, 85748 Garching, Germany}
\affil{$^{32}$ Universit\"ats-Sternwarte, Fakult\"at f\"ur Physik, Ludwig-Maximilians Universit\"at M\"unchen, Scheinerstr. 1, 81679 M\"unchen, Germany}
\affil{$^{33}$ Harvard-Smithsonian Center for Astrophysics, Cambridge, MA 02138, USA}
\affil{$^{34}$ Australian Astronomical Optics, Macquarie University, North Ryde, NSW 2113, Australia}
\affil{$^{35}$ Instituci\'o Catalana de Recerca i Estudis Avan\c{c}ats, E-08010 Barcelona, Spain}
\affil{$^{36}$ Department of Astrophysical Sciences, Princeton University, Peyton Hall, Princeton, NJ 08544, USA}
\affil{$^{37}$ Instituto de F\'\i sica, UFRGS, Caixa Postal 15051, Porto Alegre, RS - 91501-970, Brazil}
\affil{$^{38}$ School of Physics and Astronomy, University of Southampton,  Southampton, SO17 1BJ, UK}
\affil{$^{39}$ Instituto de F\'isica Gleb Wataghin, Universidade Estadual de Campinas, 13083-859, Campinas, SP, Brazil}
\affil{$^{40}$ Computer Science and Mathematics Division, Oak Ridge National Laboratory, Oak Ridge, TN 37831}
\affil{$^{41}$ Institute of Cosmology and Gravitation, University of Portsmouth, Portsmouth, PO1 3FX, UK}

\begin{abstract}
Since first noticed by Shapley in 1939, a faint object coincident with the Fornax dwarf spheroidal has long been discussed as a possible sixth globular cluster system. However, debate has continued over whether this overdensity is a statistical artifact or a blended galaxy group.
In this \textit{Letter} we demonstrate, using deep DECam imaging data, that this object is well resolved into stars and is a \textit{bona fide} star cluster. The stellar overdensity of this cluster is statistically significant at the level of $\sim$ 6 - 6.7 $\sigma$ in several different photometric catalogs including {\it Gaia}. Therefore, it is highly unlikely to be caused by random fluctuation. We show that Fornax 6 is a star cluster with a peculiarly low surface brightness and irregular shape, which may indicate a strong tidal influence from its host galaxy. The Hess diagram of Fornax 6 is largely consistent with that of Fornax field stars, but it appears to be slightly bluer. However, it is still likely more metal-rich than most of the globular clusters in the system. Faint clusters like Fornax 6 that orbit and potentially get disrupted in the centers of dwarf galaxies can prove crucial for constraining the dark matter distribution in Milky Way satellites.
\end{abstract}


\section{INTRODUCTION}
\label{intro}
The first globular cluster (GC) around the Milky Way (MW) was discovered by Jonathan Ihle in 1655. Since then, the sample size of MW GCs has grown to more than a hundred \citep{Harris_1996,Harris_2010}. A particular noteworthy group among those are clusters that orbit in the MW dwarf satellite galaxies. All three of the most luminous MW satellites, the Large and Small Magellanic Cloud, and the Sagittarius dwarf spheroidal (dSph) galaxy have large populations of well-studied GCs \citep{Mackey_etal2003a,Mackey_etal2003b,Mackey_etal2003c,McLaughlin_etal2005}. The Fornax dSph galaxy, which is the fourth most luminous MW satellite, is known to possess five GCs. Up until recently, when a peculiar faint cluster was found in an ultra-faint system, Eridanus~II \citep{Koposov_etal2015,Bechtol_etal2015,Crnojevic_etal2016}, those were the only four known MW satellites with GCs. 

Among those GC systems, clusters in the Fornax dSph are particularly interesting. The fact that Fornax has at least five of them already implies high GC specific frequency or ratio of GC mass versus halo mass, which is atypical among dwarf galaxies of similar luminosity \citep{Harris_etal2017}. The existence of this handful of GCs also presents a puzzle for understanding their survival over a Hubble time \citep{Goerdt_etal2006, Penarrubia_etal2009, Cole_etal2012,Boldrini_etal2018}. The mechanism of dynamical friction can cause those clusters to sink to the Fornax center and/or be disrupted. This so-called ``Fornax timing problem'' provides powerful constraints on the inner dark matter distribution of the Fornax dSph.

Despite Fornax's already unusually high number of GCs, there is a long and often forgotten debate in the literature about a possible sixth GC, named Fornax~6. First mentioned by Shapley in 1939 \citep{Shapley_1939}, it was later observed in greater detail by a few groups in the 80's and 90's \citep{Verner_etal1981,Stetson_etal1998,Demers_etal1994}. However, it was then thought to be a mixture of distant galaxies and stars. In this {\it Letter} we utilize several datasets to investigate the properties of Fornax~6 and demonstrate that it is a diffuse but {\it bona fide} star cluster that is likely undergoing tidal disruption.

\begin{figure}
\centering
\includegraphics[height=8.6 cm]{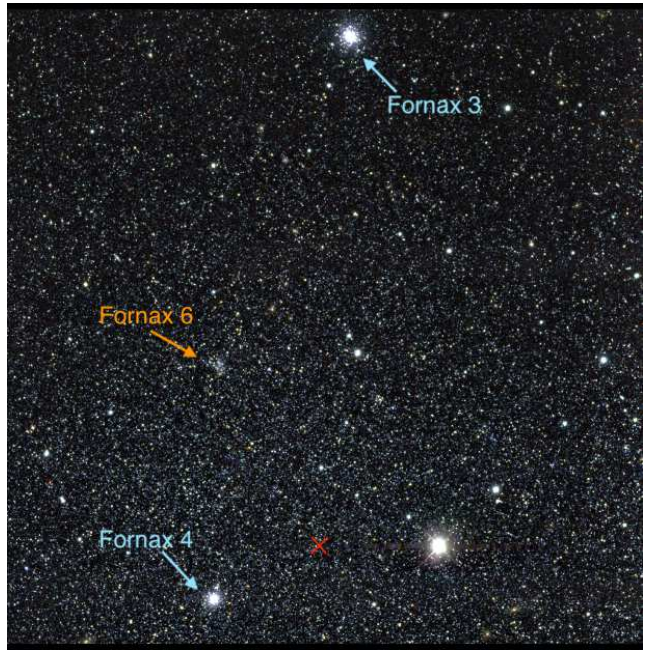}
\caption{False color $\textit{gri}$ coadded image of $\sim$ 20${\mathrm '} \times$ 20${\mathrm '}$ field around center area of the Fornax dSph galaxy from DES. The locations of several Fornax GCs including Fornax 6 are marked by blue and orange arrows. The red cross marks the center of the Fornax dSph galaxy from \cite{Wang_etal2019}.}
\label{fig:GC_location}
\end{figure}


\begin{figure*}
\centering
\includegraphics[height=7.5 cm]{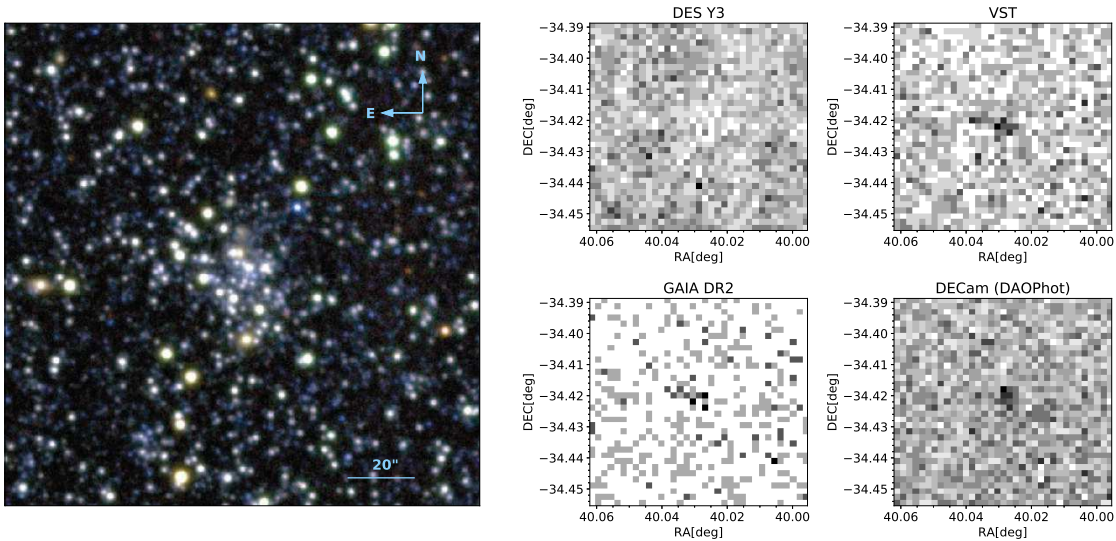}
\vspace{-1.0em}
\caption{\textit{Left panel}: False color $\textit{gri}$ coadded image of 2${\mathrm '} \times$ 2${\mathrm '}$ field centered on Fornax 6 from DES.
\textit{Right four panels}: Density distribution of detected sources around Fornax 6 in a 4${\mathrm '} \times$ 4${\mathrm '}$ field. From left to right, top to bottom, we show data from internal DES Y3, VST ATLAS, {\it Gaia} DR2, and our DECam \code{DAOPHOT} catalogs.}
\label{fig:density_dis}
\end{figure*}

\section{DATA}
\label{data}
Fornax~6 is a compact stellar overdensity in the projected central area of the Fornax dSph (see Fig.~\ref{fig:density_dis}) with no association with the system's known GCs. It was noticed by the authors during visual inspections of the DES Y3 coadded images. Because of crowding issues in the DES Y3 image processing at the center of Fornax dSph \citep{Wang_etal2019}, most of the objects in the overdensity are missing in the internal DES Y3 and the public DES DR1 source catalogs \citep{Morganson_etal18,DES_DR1}. Although the rediscovery of the overdensity was done using DES imaging, we searched for best quality publicly available data that has covered the same area. We found that the area around Fornax 6 was observed in $\textit{r}$ and $\textit{i}$ bands during exceptional seeing conditions of 0${\arcsec}$.6-0${\arcsec}$.7 by a DECam program 2016B-0244 (PI: B. Tucker) that is available in the NOAO Science Archive. We therefore use the photometrically and astrometrically calibrated catalog generated by running the DECam community pipeline \citep{valdes2014} on this imaging data for the remainder of the paper. We stack $\sim 6$ $\textit{r}$ and $\textit{i}$ band images using \code{SWARP} software \citep{Bertin2010} and calibrate the photometry by computing zero-points from the cross-match in the DES DR1 catalog. To alleviate crowding issues we ran \code{DAOPHOT} \citep{Stetson_1987,photutils} software on a region of 6\arcmin$\times$ 6\arcmin around the overdensity. 

We also utilize {\it Gaia} Data Release 2 (DR2) data \citep{Gaia_DR2} for additional photometry and proper motion information. In order to reject extended sources, we adopt the following magnitude-dependent cut on the \code{astrometric\_excess\_noise} (AEN) parameter \citep{Koposov17}:
\begin{equation}
{\rm log_{10} (\code{AEN}) <1.5+0.3(G-18)}
\label{eq:astrometric_cut}
\end{equation}

Additionally we use the data from the VST ATLAS survey \citep{Shanks15} that was reprocessed and calibrated by \citet{Koposov2014}, and spectroscopic data from \citet{Walker_etal2009}. 

\section{Objects in Fornax 6}
\label{objects}

In Figure~\ref{fig:GC_location} we show the false color $\textit{gri}$ coadded image of $\sim$ 20${\mathrm '} \times$ 20${\mathrm '}$ field around center area of the Fornax dSph galaxy from DES. The position of Fornax~6 (marked by an orange arrow) is $\sim$ 7\arcmin\ north from Fornax~4 (another Fornax GC, marked by a blue arrow) as described in the literature \citep[e.g.,][]{Shapley_1939}. It is the second closest cluster to the Fornax center in terms of projected distance ($\sim$ 0.27\,kpc) other than Fornax~4 ($\sim$ 0.15\,kpc). The Fornax dSph optical center location adopted here is from recent photometry studies \citep[e.g.,][]{Wang_etal2019,Bate_etal15}.


However, its luminosity and morphology seems peculiar among the known Fornax GCs, as it appears visually much fainter than other clusters. Also, while GCs are typically spherical, Fornax 6 has a non-negligible ellipticity (e.g., see Figure~\ref{fig:GC_location} and Figure~\ref{fig:density_dis}). In the left panel of Figure~\ref{fig:density_dis} we show the false color $\textit{gri}$ DES coadded image of 2${\mathrm '} \times$ 2${\mathrm '}$ field centered on Fornax 6. Several literature sources \citep[e.g.,][]{Verner_etal1981,Stetson_etal1998} have considered Fornax~6 as a mixture of stars and galaxies. However, most of the objects discussed in \cite{Stetson_etal1998} that were thought to be nonstellar, are clearly multiple closely located stars in the DES coadded image. Although a few of the objects are still likely distant galaxies, their contribution is far from significant.  

In the right four panels of Figure~\ref{fig:density_dis} we show the density distributions of sources from DES Y3, VST, {\it Gaia} DR2, and the \code{DAOPHOT} catalogs constructed from the DECam images. We note that barring {\it Gaia} data we do not apply any star/galaxy separation criteria as most of the sources are expected to be stars. Fornax~6 appears as a prominent overdensity in VST, {\it Gaia}, and DECam data, with the exception of DES Y3, where it is underdense, in part due to a known limitation of catalogs generated by \code{SExtractor}\citep{Bertin_etal02} in a dense star field.

We also remark that in several literature studies, Fornax~6 is discussed as a statistical artifact caused by random clustering \citep[e.g.,][]{Demers_etal1995}. To validate the presence and significance of the overdensity, we randomly draw 10000 sub-samples from the \code{DAOPHOT} and the {\it Gaia} DR2 catalog within a search radius of 13.2$\arcsec$ in a 6$\arcmin \times 6\arcmin $ field excluding area within 33.6 arcsec (two times the project half-light radius $r_h$ derived in Section~\S~\ref{profile}) centered on Fornax 6. The search radius size of 13.2$\arcsec$ is chosen to maximize the significance. 
The average and the standard deviation of the counts from the \code{DAOPHOT} ({\it Gaia}) catalog sampling is 48.9 $\pm$ 6.6 (4.0$\pm$2.0). The Fornax~6 star number count within 13.2$\arcsec$ is 96 (16), which is 6.7 (6.0) $\sigma$ above the background assuming a Poisson distribution. We note that only 1 out of 10000 random sub-samples has significance above 4 $\sigma$ (which is 4.2 $\sigma$) in the \code{DAOPHOT} catalog. For {\it Gaia} we test eight additional fields at the same distance from the Fornax dSph center. Five out of the eight fields each has 1 out of 10000 samples with significance above 4 $\sigma$ (the highest one is 4.9 $\sigma$). Therefore we conclude that Fornax 6 is highly unlikely to be a random statistical fluctuation. 

\begin{figure*}
\centering
\includegraphics[height=9.0 cm]{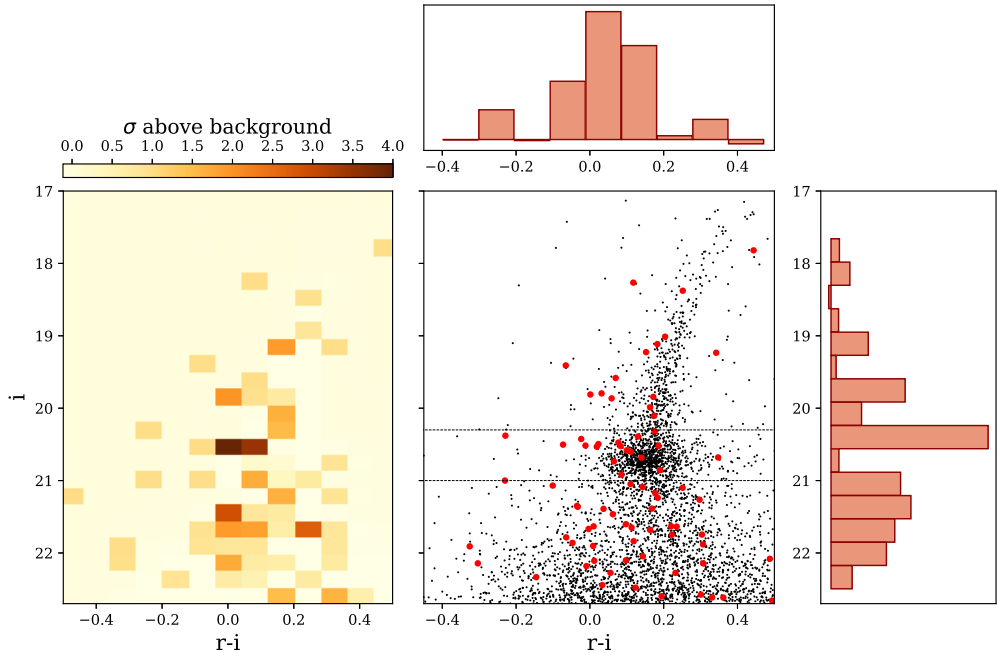}
\vspace{-1.2em}
\caption{\textit{Left panel:} Background-subtracted (with respect to Fornax dSph field stars) CMD of Fornax 6 colored by Poisson significance. \textit{Top panel:} Background-subtracted histograms of star counts as a function of \textit{r}-\textit{i} between 20.3 $<$ \textit{i} $<$ 21.0 for Fornax 6.
\textit{Middle panel:} Comparison of Fornax 6 CMD within $r_h=16.8\arcsec$(red points) with Fornax dSph CMD (black points). 
\textit{Right panel:} Background-subtracted histograms of star counts as a function of \textit{i} magnitude for Fornax 6.}
\label{fig:CMD_hist}
\end{figure*}


\section{Properties of Fornax 6}
\label{properties}

\subsection{Hess diagram}
\label{CMD}
In the middle panel of Fig.~\ref{fig:CMD_hist} we show the extinction-corrected color-magnitude diagram (CMD) of Fornax 6 (red points) within 16.8$\arcsec$ (the size of $r_h$) of the best-fit centroid from the DECam \code{DAOPHOT} catalog. The CMD of the Fornax dSph, drawn from a 5$\arcmin \times 5\arcmin$ field centered on Fornax 6, while excluding the region within 2 $r_h$ of the cluster center, is shown with black dots. In general, the CMD of Fornax 6 is largely consistent with the Fornax dSph CMD. 

The background-subtracted luminosity function (with respect to the Fornax dSph field stars) of Fornax 6, which is shown in terms of star count histograms on the right most panel in Fig.~\ref{fig:CMD_hist}, peaks at similar magnitude (i $\sim$ 20.5) as the Fornax dSph. However, this peak in the Fornax dSph is dominated by red-clump (RC) stars while in Fornax 6 it is likely dominated by the red horizontal branch (RHB) stars. Nevertheless, the similar magnitude locations of these peaks indicate that they are at comparable distances. On the top panel of Fig.~\ref{fig:CMD_hist}, we show the background-subtracted star counts as a function of \textit{r}-\textit{i} color between 20.3 $<$ \textit{i} $<$ 21.0, which covers RC and horizontal branch (HB) stars. The star counts of Fornax~6 peak at slightly bluer color than Fornax dSph. In the left panel of Fig.~\ref{fig:CMD_hist} we show the background subtracted CMD of Fornax~6 colored by their significance. The subtracted CMD also shows that the RGB stars are slightly shifted to the bluer end with additional significant features at $\sim$ 4 $\sigma$ at \textit{r}-\textit{i} = 0.0 and $i = 20.4$, likely caused by the RHB stars. This indicates that Fornax 6 may have a more metal-poor population than the inner Fornax field stars \cite[with mean ${\rm [Fe/H] \sim}$ -0.9,][]{Battaglia_etal06}. However, Fornax~6 is likely to have similar metallicity as Fornax 4 (with ${\rm [Fe/H]} \sim -1.5$) which also presents a prominent population of RHB stars and lacks a significant blue HB population unlike other Fornax GCs \citep{deBoer_etal2016}. 

\subsection{Structural Properties and Luminosity}
\label{profile}
\begin{figure}
\centering
\includegraphics[height=6.1cm]{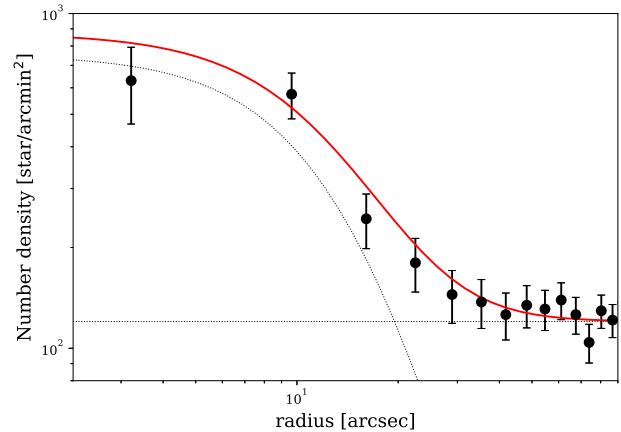}
\caption{The 1D density profile of Fornax 6. The black dash lines show the fitted Plummer and background model separately, and the red line shows their combination. Black points show the binned \code{DAOPHOT} data.}
\label{fig:1D_profile}
\end{figure}

We perform a Plummer model fit to the Fornax 6 stellar distribution using the 2D unbinned maximum likelihood algorithm described in \cite{Martin_etal2008}. We include in the fits only stars with r $<$ 22 to reduce incompleteness.   
We find that Fornax 6' s stellar density profile is well described by a Plummer profile (see Figure~\ref{fig:1D_profile} ) with $r_h$ = 16.8${\arcsec}$ $\pm$ 2.0$\arcsec$, and it has high ellipticity ($\epsilon$ = 0.41 $\pm$ 0.10). 

We use the posterior distributions of the structural parameters to estimate the total number of stars in the cluster. Then, the number of stars is converted into the total luminosity assuming a PARSEC isochrone model \citep{Bressan_etal2012} with age = 10 Gyr and [Fe/H] = -1.5 and the Chabrier initial mass function (IMF) \citep{Chabrier_2003}. Assuming a distance of 147 kpc \citep[the distance of Fornax dSph,][]{McConnachie_12}, the estimated absolute magnitude of Fornax 6 in V band is ${\rm M_V}=-4.8 \pm 0.4$, which is fainter than other Fornax GCs that range from ${\rm M_V}=-8.2$ to $-5.2$ \citep{Webbink_1985}. A summary of Fornax 6 properties is provided in Table~\ref{tb:properties}.

\subsection{Radial and Proper Motion Velocity}
\label{vel}

To test the hypothesis of Fornax 6 belonging to the Fornax dSph, we use {\it Gaia} to investigate what fraction of stars in Fornax 6 are bound to the Fornax dSph. To identify possible members, we select stars within 2 $r_h$ of the cluster center in {\it Gaia} and require that stars have small parallax with respect to its uncertainties of $\varpi < 2\, \sigma_\varpi$. We also require the proper motion to be within 3 $\sigma$ from the estimated escape speed at the distance of the object. In addition, we apply a CMD mask using {\it Gaia} photometry to select targets lying within $\pm$ 0.2 mag to the isochrone model mentioned in previous section. We find that 18 out of 32 (56$\%$) of stars lying within 2$r_h$ of Fornax 6 pass these criteria, while it is expected that only 8 stars ($\sim$ 25$\%$) should come from Fornax dSph field star contamination. Therefore a significant fraction of Fornax 6 stars has proper motion consistent with being bound to the Fornax dSph.

\begin{table}
\caption{Properties of Fornax 6}
{\renewcommand{\arraystretch}{1.5}
\renewcommand{\tabcolsep}{0.7cm}
\centering
\begin{tabular}{c c c }
\hline 
\hline
Parameter & \\
\hline
RA(J2000) & ${\rm 2^h40^m6.9^s}$  \\
Dec(J2000) & ${\rm -34^{\circ} 25{\mathrm '}19.2{\arcsec}}$ \\
$\epsilon$& 0.41$\pm$0.10\\
P.A. (deg) & $13.1^{+10.4}_{-7.3}$  \\
$r_h$ (arcsec) & 16.8 $\pm$ 2.0 \\
$r_h$ (pc) & $11.3 \pm 1.4$ \\
${\rm M_V}$ (mag) & -4.8$\pm$ 0.4\\
\hline
\end{tabular}\\
\vspace{0.5em}
 }
\label{tb:properties}
\end{table}

There is a small sample of 5 stars within 2.2 $r_h$ with spectroscopic measurements from \cite{Walker_etal2009}. However, only 4 of them have proper motion consistent with being bound to the Fornax dSph. The derived mean radial velocity (RV) and velocity dispersion of these 4 stars is $53.9^{+1.2}_{-1.1}$\,km/s and $2.1^{+1.7}_{-0.8}$\,km/s. Its RV is very similar to the RV of the Fornax dSph, which is $53.3$\,km/s \citep{Walker_etal2009}. We note that $\sim 30\%$ of stars in this sample are expected to be background stars belonging to Fornax dSph. However if we randomly draw 4 stars out of the velocity distribution of Fornax dSph, which has velocity dispersion of $11.7$\,km/s \citep{Walker_etal2009, McConnachie_12}, the probability of measuring dispersion less than $3.8$\,km/s (within 1$\sigma$ upper bound of our estimation) is only $\sim$ 6.5$\%$. Thus under the assumption that some of the measured signals indeed correspond to Fornax~6, it indicates that this cluster has low velocity dispersion and a RV that is similar to the Fornax dSph itself.

\section{Implications for Fornax dSph dark matter distribution}
\label{DM}
The tidal evolution of a diffuse GC like Fornax 6 may provide important clues for understanding Fornax's inner dark matter profile. For example, \cite{Penarrubia_etal2009} argues that surviving low-mass GCs like Fornax~1 should mostly populate the outskirts of their host galaxies to avoid being disrupted by tides. Furthermore, it is expected that orbital decay due to dynamical friction experienced by a low-mass GC would be small \cite[e.g.,][]{Binney_etal08}. Thus Fornax 6' s small projected distance to the center of Fornax is in tension with these arguments, although its true 3D distance is unknown. Therefore there are two possible scenarios: (1) the initial mass of Fornax 6 was large and comparable to Fornax 2 - 5, and therefore its 3D distance to the Fornax dSph center could be close due to significant orbital decay. It also implies substantial mass loss. (2) The initial mass of Fornax 6 was small, and therefore its 3D distance may actually not be very close to the center of Fornax.  

Here we estimate the influences of tides by computing the Jacobi (tidal) radius $r_J$. The average stellar density of the cluster within $r_h$ is $\sim$0.7 ${\rm M_\odot/pc^3}$. At $r=0.27$\,kpc (the projected distance to Fornax center), the Fornax dSph dark matter density $\rho_{dm}$ is $\sim$ 0.08 ${\rm M_\odot/pc^3}$ for a NFW potential with $V_{\rm max} \sim$ $30$\,km/s. Therefore the $r_J$ for Fornax~6  is $\sim 16$\,pc. This is only factor of 1.4 times the cluster half-light radius, and therefore supports the tidal disruption hypothesis. If we assume a cored profile, such as the one from \citep{Walker_etal2011} (see their Figure 1) that suggests $\rho_{dm}$ is $\sim$ 0.04 ${\rm M_\odot/pc^3}$ at $r=0.27$\,kpc; the predicted $r_J$ is $\sim 21$\,pc (1.9 times the half-light radius). We also remark that Fornax 6 could be strongly affected by tides in a NFW potential even if it is significantly farther away from the Fornax center, as the $r_J$ can be smaller than $2\times r_h$ for 3D separations up to 0.8 kpc. In contrast, in the cored profile, the $r_J$ stays $\sim$ 1.8-2.0$\times r_h$ anywhere within a 3D distance of 0.8 kpc, indicating limited tidal influences when the inner host potential is shallow. 

Due to its small mass, this rediscovered star cluster does not add much to the total GC mass budget associated with the Fornax dSph, and therefore does not impact on the established GC mass -- halo mass relation. However, the total GC mass budget within a galaxy may have been underestimated due to substantial amount of tidally stripped stars buried in the galaxy's dense stellar field.

\section{Conclusion}
\label{conclusion}

In this \textit{Letter} we demonstrate that  Fornax~6, which was historically thought to be a dubious object, is a genuine star cluster within the Fornax dSph. Using deep DES Y3 coadded images we show that a few objects within Fornax 6 that are claimed to be galaxies in the literature are actually blended images of multiple stars. The stellar overdensity caused by Fornax 6 has a significance of  6-6.7 $\sigma$ in the DECam and the {\it Gaia} DR2 photometry catalogs.  We show that Fornax 6's  CMD is largely consistent with Fornax dSph, but slightly shifted to the bluer end. 

Fornax~6' s light profile is well-fit by a Plummer model with $r_h$ = $11.3 \pm 1.4$\,pc and high ellipticity of $\epsilon = 0.41 \pm 0.1$, with an estimated luminosity of $M_V$ = -4.8$\pm$0.4. The highly elongated shape of Fornax~6 suggests that it is undergoing tidal disruption. 

We also check the available kinematic information to assess how likely Fornax~6 is bound to the Fornax dSph. By applying stringent selection criteria on stars in {\it Gaia}, we show that a high fraction of stars within Fornax~6 have proper motion consistent with being bound to the Fornax dSph. Four possible members of Fornax~6 with spectroscopic data also suggest radial velocity close to the velocity of the Fornax dSph with low velocity dispersion, but this needs to be verified with more members.


Since Fornax 6 may be the only Fornax GC that shows clear signs of tidal disruption, its tidal evolution can provide a powerful probe to the Fornax dSph dark matter potential. Depending on the assumptions about the initial mass of Fornax~6 and its 3-D distance to the Fornax center, the dynamical friction and tidal force it experienced can vary. Detailed N-body simulations will be needed to quantify these effects. 

\acknowledgments

\textit{Acknowledgements}: MYW acknowledges support of the McWilliams Postdoctoral Fellowship. SK is partially supported by National Science Foundation grant AST-1813881.

This paper has gone through internal review by the DES collaboration.

Funding for the DES Projects has been provided by the DOE and NSF(USA), MEC/MICINN/MINECO(Spain), STFC(UK), HEFCE(UK). NCSA(UIUC), KICP(U. Chicago), CCAPP(Ohio State), 
MIFPA(Texas A\&M), CNPQ, FAPERJ, FINEP (Brazil), DFG(Germany) and the Collaborating Institutions in the Dark Energy Survey.

The Collaborating Institutions are Argonne Lab, UC Santa Cruz, University of Cambridge, CIEMAT-Madrid, University of Chicago, University College London, DES-Brazil Consortium, University of Edinburgh, ETH Z{\"u}rich, Fermilab, University of Illinois, ICE (IEEC-CSIC), IFAE Barcelona, Lawrence Berkeley Lab, LMU M{\"u}nchen and the associated Excellence Cluster Universe, University of Michigan, NOAO, University of Nottingham, Ohio State University, University of Pennsylvania, University of Portsmouth, SLAC National Lab, Stanford University, University of Sussex, Texas A\&M University, and the OzDES Membership Consortium.

Based in part on observations at Cerro Tololo Inter-American Observatory, National Optical Astronomy Observatory, which is operated by the Association of Universities for Research in Astronomy (AURA) under a cooperative agreement with the National Science Foundation.

The DES Data Management System is supported by the NSF under Grant Numbers AST-1138766 and AST-1536171. 
The DES participants from Spanish institutions are partially supported by MINECO under grants AYA2015-71825, ESP2015-66861, FPA2015-68048, SEV-2016-0588, SEV-2016-0597, and MDM-2015-0509, some of which include ERDF funds from the European Union. IFAE is partially funded by the CERCA program of the Generalitat de Catalunya.
Research leading to these results has received funding from the European Research
Council under the European Union's Seventh Framework Program (FP7/2007-2013) including ERC grant agreements 240672, 291329, and 306478.
We acknowledge support from the Australian Research Council Centre of Excellence for All-sky Astrophysics (CAASTRO), through project number CE110001020, and the Brazilian Instituto Nacional de Ci\^enciae Tecnologia (INCT) e-Universe (CNPq grant 465376/2014-2).

This manuscript has been authored by Fermi Research Alliance, LLC under Contract No. DE-AC02-07CH11359 with the U.S. Department of Energy, Office of Science, Office of High Energy Physics. The United States Government retains and the publisher, by accepting the article for publication, acknowledges that the United States Government retains a non-exclusive, paid-up, irrevocable, world-wide license to publish or reproduce the published form of this manuscript, or allow others to do so, for United States Government purposes.



\bibliographystyle{apj}

\end{document}